\documentclass[conference]{IEEEtran}
\IEEEoverridecommandlockouts
\usepackage{cite}
\usepackage{amsmath,amssymb,amsfonts}
\usepackage{algorithmic}
\usepackage{graphicx}
\usepackage{textcomp}
\usepackage{xcolor}
\def\BibTeX{{\rm B\kern-.05em{\sc i\kern-.025em b}\kern-.08em
    T\kern-.1667em\lower.7ex\hbox{E}\kern-.125emX}}
\begin{document}

\title{Web Phishing Net (WPN): A scalable machine learning approach for real-time phishing campaign detection\\
}

\author{\IEEEauthorblockN{Muhammad Fahad Zia}
\IEEEauthorblockA{\textit{Future Cyber Defence, Research and Network Strategy} \\
\textit{BT Group}\\
Martlesham Heath, UK \\
muhammadfahad.zia@bt.com}
\and
\IEEEauthorblockN{Sri Harish Kalidass}
\IEEEauthorblockA{\textit{Future Cyber Defence, Research and Network Strategy} \\
\textit{BT Group}\\
Martlesham Heath, UK \\
sriharish.kalidass@bt.com}
}

\maketitle

\begin{abstract}
Phishing is the most prevalent type of cyber-attack today and is recognized as the leading source of data breaches with significant consequences for both individuals and corporations. Web-based phishing attacks are the most frequent with vectors such as social media posts and emails containing links to phishing URLs that once clicked on render host systems vulnerable to more sinister attacks. Research efforts to detect phishing URLs have involved the use of supervised learning techniques that use large amounts of data to train models and have high computational requirements. They also involve analysis of features derived from vectors including email contents thus affecting user privacy. Additionally, they suffer from a lack of resilience against evolution of threats especially with the advent of generative AI techniques to bypass these systems as with AI-generated phishing URLs. Unsupervised methods such as clustering techniques have also been used in phishing detection in the past, however, they are at times unscalable due to the use of pair-wise comparisons. They also lack high detection rates while detecting phishing campaigns. In this paper, we propose an unsupervised learning approach that is not only fast but scalable, as it does not involve pair-wise comparisons. It is able to detect entire campaigns at a time with a high detection rate while preserving user privacy; this includes the recent surge of campaigns with targeted phishing URLs generated by malicious entities using generative AI techniques.
\end{abstract}

\begin{IEEEkeywords}
phishing detection, campaign detection, unsupervised learning
\end{IEEEkeywords}

\section{Introduction \label{intro}}

Phishing is a type of cyber-attack in which the attacker mimics a legitimate website to extract sensitive information from victims; information such usernames, emails, passwords and details of bank or credit cards. It is the most common form of cyber-crime, with an estimated 3.4 billion instances of attack every day \cite{b1}. Over 48\% of emails sent in 2022 were classified as spam with an average cost of a resulting data breach against organisations being £4 million \cite{b1}. The phishing vectors were initially web-based communications such as emails, however, there has been a change in the trend where texts, phone call and social media applications are being used as well. Nevertheless, web-based phishing remains the biggest threat in the cyber-crime domain.

Since phishing became popular in the late 2000s, there have been different techniques and methods evolved. Email phishing, spear phishing and whaling accounted for most of the phishing attacks. The key difference between these types of attacks is that email phishing scammers try to trick users into giving away personal information, whereas in spear phishing, scammers use specific information about users to make the attack seem more legitimate and trick them into an action leading to data breaches. Whaling is a also a targeted attack, however it involves high-ranking individuals as victims - CEOs or directors of corporations with access to company’s sensitive information. Business email compromise \cite{b2} and social media phishing \cite{b3} have specifically increased over the last several years. The former type of phishing targets businesses; scammers impersonate executives or employees in emails, aiming to defraud the company or its partners. The latter involves fraudsters who create fake profiles on social media or imitate seemingly legitimate posts to trick users into giving away personal information. 

Majority of all of these phishing attacks attempt to take victims to a website through phishing URLs via emails where they are tricked into giving away sensitive information. Surprisingly, only 9\% of all phishing emails are attempting to infect the victim with malware \cite{b4}.

Most of these web-based phishing attempts have a detectable pattern such as a mismatch in the sender’s email address and domain name, grammatical errors and an unusual sense of urgency or time limit. Email security solutions are designed to detect these patterns; text analysis is used to read the email header and the email itself to identify poor grammar and patterns commonly associated with phishing attacks. These patterns include use of suspicious keywords, phrases, or links that may be indicators of a phishing attempt. To detect these, top  phishing detectors widely used around the globe thus employ (in addition to header filters and blacklist filters) content filtering algorithms that analyse email content to identify keywords or patterns associated with phishing emails. However, these phishing detectors breach user privacy as they sift through sensitive and private email content to generate features needed to detect phishing.

Another shortcoming of these detection systems is their inability to efficiently detect AI-enabled attacks using Large Language Models (LLMs) like ChatGPT; models that can generate URLs/emails \cite{b5}. According to security provider Egress, 71\% of email attacks created through AI go undetected \cite{b6}. The attackers employ red versions of ChatGPT-like models; these rogue AI versions, the likes of Fraud GPT and Worm GPT, generate thousands of highly convincing phishing emails in various languages - emails containing phishing URLs which in just a few seconds can easily bypass a human firewall and even advanced traditional security systems. 

The need of the hour is thus a privacy-enhanced phishing detection solution that is able to protect users from the ever-evolving and sophisticated threat landscape of next generation phishing attacks that employ advanced technologies like Generative AI to produce attack vectors on large scales. In this paper, we present a solution to deal with the aforementioned gaps identified in current phishing detection systems. The rest of this paper is structured as follows: Section \ref{related} discusses the related work in detail. Section \ref{WPN} unravels our proposed approach and the various stages of detection. Section \ref{evaluation} covers the evaluation of our approach and experimentation; Section \ref{conclusion} covers the conclusion and future work.

\section{Related Work \label{related}}

Phishing detection systems generally employ a number of common techniques in combination to enable reliable protection against the attacks; this includes URL scanning, email content scanning, network metadata analysis (DNS blacklisting), rule-based methods and more recently AI-based phishing detection tools which use large amounts of data to train machine learning (ML) models to extract features from email contents, web-site characteristics, user interactions and other associated metadata \cite{b7}. Broadly categorizing these systems, there are primarily two types - traditional and non-traditional methods. The traditional methods primarily incorporate legal protection, user education and awareness, blacklist/whitelist, visual similarity, and search engines. The non-traditional methods fundamentally include content-based, heuristics-based, AI, fuzzy logic, rule-based and data mining methods. The proposed work falls under the latter category of AI-enabled phishing detection systems.

In the latest research involving the aforementioned category, there has been greater focus on supervised learning based machine learning techniques. A phishing URL classification using recurrent neural networks is proposed by \cite{b8}, where an LSTM model is trained on 2 million URLs containing legitimate and phishing URLs and each character of a URL input is translated to 128-dimensional embedding for training the model. This model trained on URLs was bench-marked against a random forest model trained on lexical and statistical features associated with URLs and although the Random Forest model yielded an overall accuracy of 94\%, the LSTM outperformed it by using just one feature, the URL string, yielding 99\% overall accuracy. This highlights the importance of analysing URL text in phishing detection.

The Deep Neural Network-based approach taken by \cite{b9} is another method where a large dataset consisting of malicious and benign URLs was used; in this case, to train 3 different types of Convolutional Neural Networks (CNNs). The first CNN model utilises character-level URL embeddings, the second model uses word-level embedding of the same URL and the third (and most effective) model used a hybrid approach where the embeddings used for training the CNN model uses word and character-level embeddings together.

The supervised learning approach has proven quite effective for detecting traditional phishing URLs, however, supervised machine learning are not generalizable in the face of the ever-evolving phishing URLs especially with the need for large amounts of labelled data and feature engineering processes, which are prone to human error \cite{b11}. Additionally, classical machine learning techniques still suffer from a lack of efficiency in detecting zero-day phishing attacks \cite{b12}. Consequently, supervised machine learning methods fail to detect unknown or newly evolved phishing attacks. Deep learning algorithms such as CNNs can detect zero-day attacks more efficiently \cite{b13}. However, although these approaches are still dependent on large amounts of historical labelled data for training, and are thus not as effective against the evolving techniques of attack and not resilient to generative AI-based techniques. Prior to \cite{b10}, most of the phishing URL detection research was focused on supervised classification techniques, and there is very limited work done on adapting clustering for efficient phishing URL detection.

The work published in \cite{b10} emphasises on the importance of using unsupervised methods such as clustering, in contrast to previously cited works which employ supervised algorithms requiring computation resources and large amounts of data to train models. It explores clustering methods such as K-Means and DBSCAN for detecting phishing URLs. The workflow starts with the browser analytics segment which monitors the user’s activity and any domain searched is matched with white and blacklists. If there is a match with blacklist the domain is blocked and vice-versa. If there are no matches, then certain features from the website are derived, such as website protocol, IP address subdomains, the amount of web traffic and Google Page-rank index etc. Based on the extracted features clustering is carried out using K-Means clustering algorithm to output two clusters (benign and malicious). Unlike the above supervised methods, here the analysis for classifying a URL is done by analysing the external features in addition to the URL itself. A shortcoming of these particular unsupervised approaches is their high computation requirements as data volumes increase and thus lack of scalability in addition to lack of accuracy with high-dimensional data such as text \cite{b22}.

There is a need for further research into evolving web-based phishing detection approaches; more specifically into the unsupervised clustering techniques that hold the potential for detection of evolved threats (without the need for historical training data). The proposed method in this paper aims to fill that gap and offers a solution using unsupervised techniques, which not only addresses the aforementioned shortcomings but also enables scalability and resilience, traits which are unique to this method. Applications of our approach include but are not limited to detection of entire phishing campaigns using a scalable and efficient pipeline, enabling detection of zero-day attacks with detection of newly registered phishing domains (which do not offer additional features such as traffic data) and resilience against AI-generated phishing campaigns; all-the-while ensuring privacy of users’ content.

\section{Web Phishing Net (WPN) \label{WPN}}

WPN is an unsupervised phishing detection system that uses an efficient 3-stage pipeline with unsupervised hash-based clustering of data to detect and classify entire campaigns containing URLs associated with phishing. 

The system leverages the key property of phishing URLs in that they tend to mimic known legitimate domains to entrap victims. This allows the system to detecting phishing URLs by clustering new observations of URLs together with both known legitimate domains as well as known phishing domains/URLs. If a new observation is clustered with either known legitimate domain or a known phishing URLs, due to its similarity to them, it is classified as a phishing URL and following verification fed back into the pool of known phishing URLs to ensure the system uses the feedback loop to improve overtime. 

\begin{figure}[htbp]
\centerline{\includegraphics[width=3.3in]{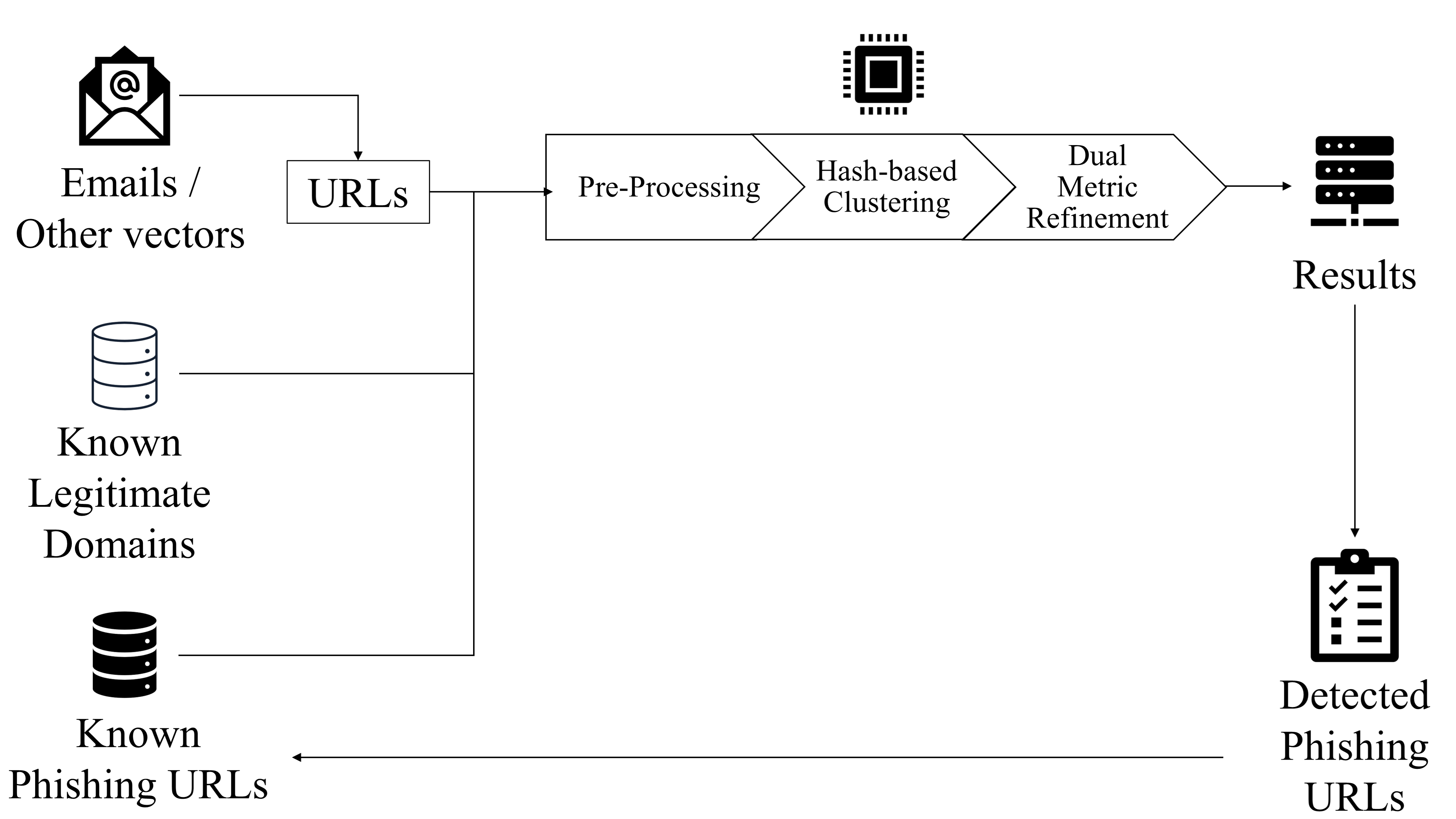}}
\caption{Architecture diagram of WPN solution for phishing use case.}
\label{fig1}
\end{figure}

The flowchart in Fig. \ref{fig1} illustrates this approach. It comprises of the 3 stages: pre-processing, hash-based clustering, and dual metric refinement. 

Input data, comprising unlabelled domains, known legitimate domains and known phishing URLs, pass through the first pre-processing phase; this cleans up the strings in preparation for clustering which requires tokenized inputs as numerical embedding vectors. The second clustering phase then uses a hashing-based clustering algorithm to efficiently bin or cluster similar domains/URLs together; this clusters the unlabelled URLs that are to be detected with known legitimate domains and known phishing URLs, allowing the system to detect entire phishing campaigns without the need for pair-wise comparisons. Unlabelled inputs that are part of clusters with known legitimate domains or known phishing URLs based on their similarity to them are thus marked as phishing URLs. The final phase then post-processes the clusters to ensure false positives are removed by using a combination of similarity measures and thresholding to remove non-phishing URLs from the final result set. Below are the details of the 3 stages in detail.

\subsection{Pre-processing}

\begin{figure}[htbp]
\centerline{\includegraphics[width=3.2in]{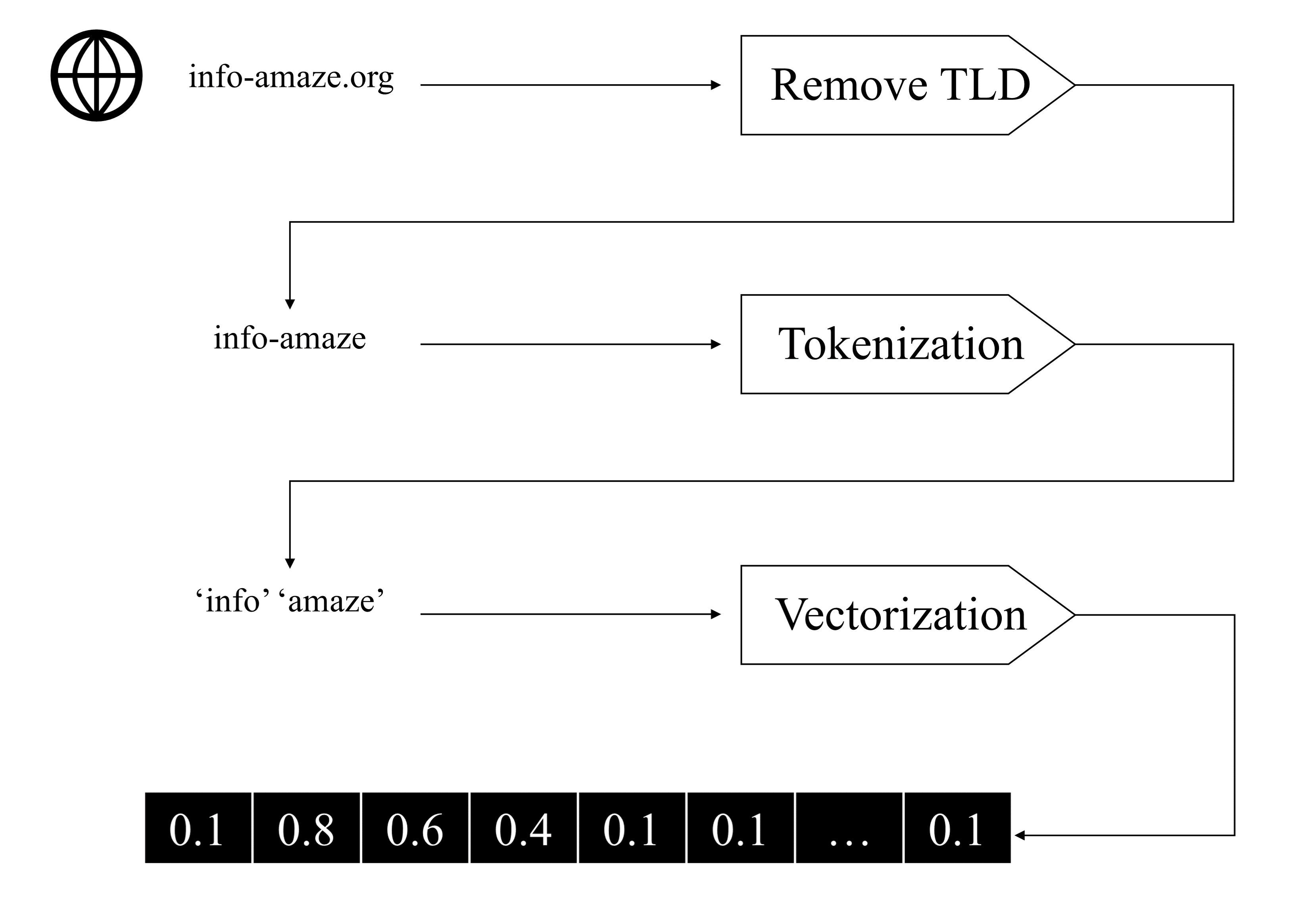}}
\caption{Working example of WPN pre-processing stage.}
\label{fig2}
\end{figure}

The pre-processing step takes URLs strings as input, and outputs vector representations that capture the lexical features of these URLs. The procedure is illustrated in Fig. \ref{fig2} and involves three steps. The first is removing the top-level domain (TLD); this is done to be able to capture the similarity between domains, regardless of the difference in their TLDs. Additionally, given that TLDs are a finite set, it is not useful to boost the detected similarity between domains/URLs simply because their TLDs match. Once TLDs have been removed, the next step is to break up the string into individual tokens (words); this is the tokenization process. All tokens extracted from the input dataset are collected together into a set to form a “vocabulary”. Each input string can thus be represented as a vector with dimension equal to the size of the vocabulary, where all tokens present in the URL have a non-zero entry in the vector. This numerical representation in the form of vectors is then passed to the following step.

\subsection{Hash-based Clustering}

This step essentially groups a set of items such that items in the same cluster (group) are more similar to each other than to items in another cluster. Through this step, we cluster input (unlabelled) URLs with known legitimate URLs and known phishing ones, such that items that are grouped together in the same cluster are similar enough for the input URLs in the cluster to be classified as phishing. This similarity is generally calculated using on the vector representations from the pre-processing step using geometrical measures such as Euclidean distance or cosine similarity.

There are several clustering algorithms that can be deployed to achieve this purpose. A popular clustering algorithm, K-means \cite{b14}, is designed to group items in one of k (k\textgreater1) clusters based on the proximity of the items to k centroids (or anchor points). K-means algorithm is not selected for our approach we aim for URLs/domains to be put in the same cluster only if they are highly similar i.e., similarity between the domains is above a threshold at which they can be judged as being similar enough to be likely part of the same campaign. This is opposed to the K-Means algorithm whether clustering is done based on a pre-decided number of clusters. In our application of phishing, we are not able to determine the number of clusters, k, in advance. Additionally, unlike in K-Means, if no subset of domains is similar enough, we want no clusters to be returned by the algorithm. In a real-world setting, we expect only a small subset of URLs out of all the scanned inputs to be above the similarity threshold (and thus be potentially phishing URLs); most input URLs will therefore not belong in any cluster.

DBSCAN \cite{b15}, another popular clustering algorithm, identifies clusters based on the notion that a cluster in feature space is a contiguous region of high item density. For our application, as mentioned previously high density of URLs per clusters is not a requirement; rather in detecting phishing URLs, highly similar items inside clusters is a stronger objective.

BIRCH is a scalable hierarchical option in clustering algorithms; however it is limited by its operation in the Euclidean space and lack of complexity in cluster shapes. Hierarchical Agglomerative Clustering (HAC), another popular hierarchical clustering technique, starts off by treating each item as belonging to its own individual cluster and proceeds to merge most similar pairs of clusters, until all clusters have been merged into one or until a similarity threshold has been reached. The approach of HAC is quite suitable for this application as it only forms clusters from smaller clusters whose similarity is above a threshold, however, the exhaustive pairwise similarity computation between all domains makes HAC very slow and computationally expensive for our application. 

We therefore adopt a clustering approach based on Locality Sensitive Hashing (LSH) \cite{b16}. LSH  is an algorithm for assigning items into buckets using a hashing function, such that highly similar items have a high probability of being assigned to the same bucket. In other words, LSH assigns similar items to the same bucket and dissimilar items to different buckets. Although LSH was not traditionally designed for clustering, and is widely used in approximate nearest neighbour search, the buckets of similar items generated by LSH are essentially clusters of similar items. Additionally, the efficiency of this approach makes it suitable for our application as it is designed to scale to larger datasets especially those with high dimensionality \cite{b22}, as is the case with text-based inputs in our phishing scenario. 

\begin{figure}[htbp]
\centerline{\includegraphics[width=3.5in]{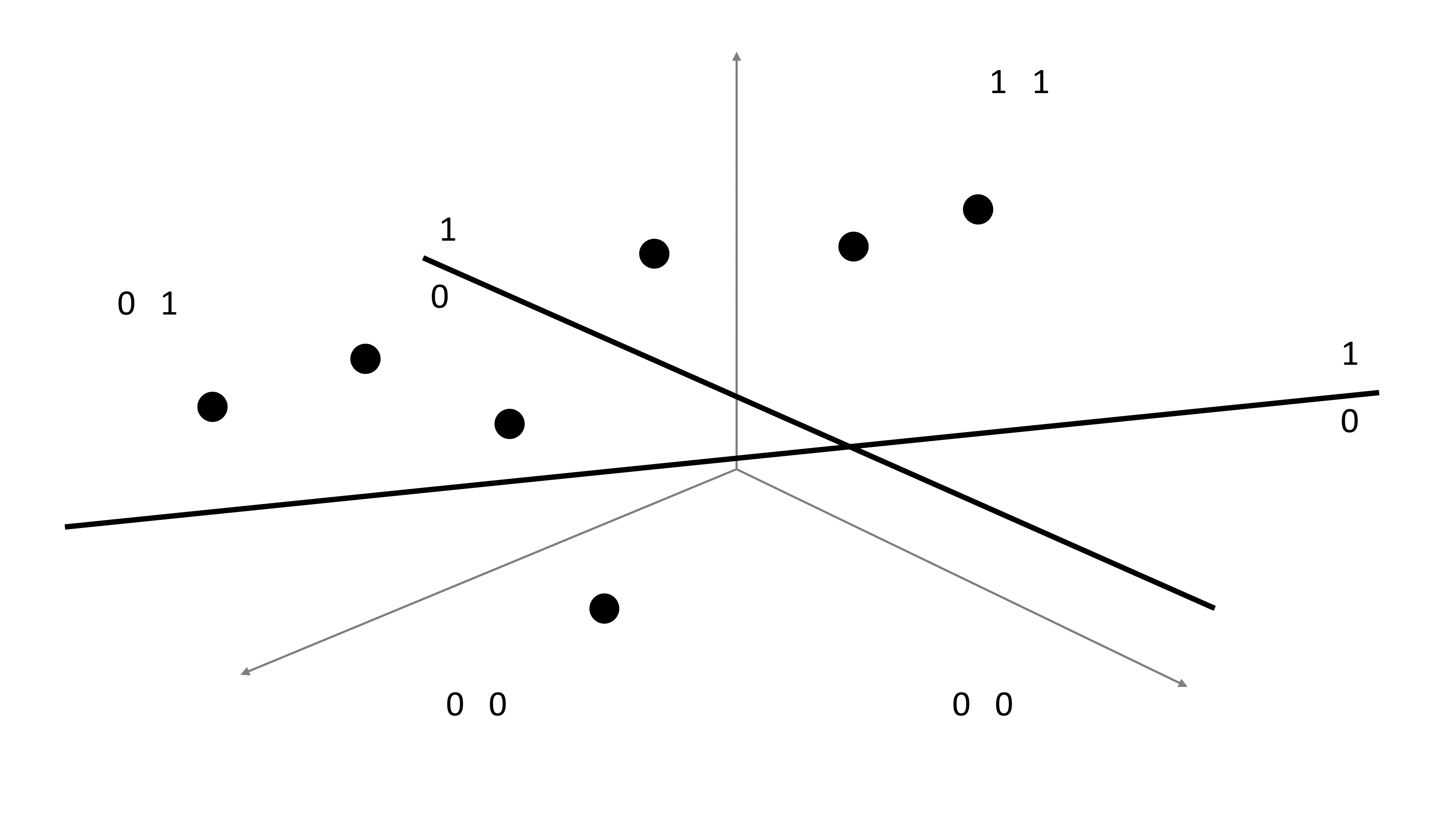}}
\caption{Illustration of LSH clustering with \textit{k}=2.}
\label{fig3}
\end{figure}

We adopt one particular variant of LSH \cite{b17} in the proposed WPN system. Given N items (URLs) to cluster with LSH, where each item is a vector in d dimensional space, we create \textit{k} random d dimensional vectors called projection vectors. The value of the parameter \textit{k} needs to be chosen and controls the number of clusters formed, and the how similar items need to be to end up in the same bucket. The projection vectors divide the space into $2^k$  possible buckets or clusters such that \textit{k}=2 generates 4 potential clusters and \textit{k}=10 generates 1024. This is a reason for the efficiency of LSH; to generate 1024 clusters, only 10 projection vectors are needed thus saving computation and space. Fig. \ref{fig3} illustrates a working of the LSH with an example \textit{k} equal to 2.

The output of the LSH results in clusters of similar domains/URLs where input phishing URLs are grouped together with known legitimate and/or known phishing URLs. This grouping allows us to detect entire campaigns of phishing URLs at a time.

\subsection{Dual Metric Refinement}

Once all URLs have been clustered, we apply a set of rules to refine the clusters to produce the final set of detections. Recall that the goal of this application is to find clusters of URLs that are highly similar to known legitimate domains or known phishing URLs (for them to be classified as phishing). Therefore, the following refinement steps are designed to ensure the output meets this goal. 

Firstly, we ensure each cluster that is shortlisted contains input phishing  URLs mixed with either known legitimate domains or known phishing URLs. Clusters that contain either only known domains/URLs or only the new input URLs are inconclusive and thus dropped from the final output as they cannot be deemed as phishing URLs with sufficient evidence. 

We then ensure all input URLs in a cluster have very high similarity to the known legitimate domains or known phishing URLs in their cluster. The previous LSH step puts similar URLs in the same cluster, however there is no strict similarity threshold imposed. At this stage, we have a chance to apply more fine-grained similarity metrics. More computationally expensive metrics are possible at this stage because the number of domains under consideration would have dropped significantly after clustering and shortlisting of clusters. Also, only URLs/domains within the same cluster are compared, thus significantly reducing the number of pair-wise comparison required. We consider two metrics (dual metric), Levenstein distance \cite{b18} and Dice distance \cite{b19} to proceed with refinement selected for their specific properties in establishing string similarity.

Levenstein distance, also called edit distance, measures the number of insertions, deletions and substitutions needed to make two strings equal. This is ideal for measuring how similar two URL strings are. However, Levenstein will give a large distance value to two otherwise identical strings whose words are put in a different order. Given that changing the order of words or \textit{spoofing} is a common technique used in phishing, an additional similarity metric that is insensitive to word order is needed to complement Levenstein. For this purpose, we combine the Levenstein distance with the Dice metric.

Dice measures the overlap in two input sets, regardless of the order of the items in the sets. Domain names can be considered as a set of tokens (obtained from the tokenization step); Dice metric allows us to measure the similarity between the two strings based on the overlap in their tokens, irrespective of order in which they occur. 

This combination of Levenstein and Dice metrics allows us to sufficiently capture the string-based similarity commonly encountered in phishing scenarios. We are then able to apply a threshold on this combination score to keep only the most similar domains/URLs in the same cluster. The combination score is calculated by taking the minimum of the two metrics and comparing this to the threshold. Example results from this process are shown in Table \ref{tab1} below with a single cluster; the top two cells of the block contain known legitimate domains while the bottom two cell contain the detected phishing URLs (from the input URLs) in the same cluster.

\begin{table}[htbp]
\caption{Sample output of final stage of WPN showing a cluster with legitimate domains and detection phishing URLs.}
\begin{center}
\begin{tabular}{|c|c|c|c|}
\hline
\textbf{URLs}&\textbf{Safe}&\textbf{Cluster}&\textbf{Common Words} \\
\cline{2-4} 
\hline
\textit{amazonlogistics.eu}&1&24&amazon  \\
\textit{creditamazon.ge}&1&24&amazon  \\
\textit{supportservice-amazon.de}&0&24&amazon  \\
\textit{kunde.supported-amazon.net}&0&24&amazon  \\
\hline
\end{tabular}
\label{tab1}
\end{center}
\end{table}

The overall approach of the WPN system has several advantages compared to other unsupervised learning-based systems; the first being the proactive or day zero detection of these domains/URLs. WPN does not require any additional data other than URL strings; this means that it can detect phishing as soon as new phishing domains are registered and URLs crop up, without waiting for IP addresses to be assigned to them or network traffic to start flowing to them. This allows proactive detection and response to reduce any harm that may come from waiting for these phishing URLs to become active. Detecting phishing URLs in this way also enables the system to detect phishing emails (that direct victims to these URLs) in a privacy-preserving manner which is in contrast with the other approaches that use content filtering techniques to sift through the email content in order to detect phishing. Additionally, the lack of pair-wise comparisons and quick binning of data using hashing enables faster processing times compared to other unsupervised algorithms and thus a scalable and efficient approach to phishing detection.

The second advantage is that it can detect entire phishing campaigns at the same time using clustering. This contrasts with alternative approaches that tend to make inferences on a single phishing URL at a time (such as aforementioned supervised techniques) or use pair-wise comparisons (such as other agglomerative clustering-based systems). Campaign detection is important because generally phishing campaigns are targeted and tend to be popular at different times. For example, the rise of cryptocurrencies led to an increase in new phishing domains with URLs using cryptocurrency keywords. Similarly, following the COVID-19 crisis, many malicious COVID-19 domains were registered with COVID-related keywords. Being able to detect entire campaigns at a time proactively reduces the potential harms from these URLs.

Finally, the approach has a lower miss rate than known unsupervised techniques (as will be evaluated in the following section) and is resilient to evolved threats which in the case of phishing can involve phishing URLs generated using generative AI and large language models (LLMs). Many traditional phishing detection approaches have been unable to defend against such evolved threats from attackers using AI for malicious activities.

\section{Evaluation \label{evaluation}}

To evaluate our approach, we conducted a series of experiments to compare each property of WPN against other popular phishing URL detection methods. As mentioned in section \ref{related}, phishing detection has seen a shift towards unsupervised detection due to its specific advantages over supervised methods which are limited by data availability and lack of generalization to evolving attacks. For the scope of this paper, therefore, we evaluated WPN against popular \textit{unsupervised} detection techniques such as those discussed in \cite{b10} which proved more favourable in phishing URL detection compared to supervised methods.

\subsection{PhishStorm Dataset}

For the first experiment, the dataset used for evaluation is the popular PhishStorm data \cite{b20} which combines a malicious URL list from PhishTank and a benign URLs list sourced from the Open Directory Project - DMOZ. The approach to curate the data leverages search engine query data in order to extract 12 features from URLs characterizing its intra relatedness and its popularity. The aim was to introduce a URL rating system, PhishStorm, to dynamically compute a risk score for URLs to classify them as phishing or legitimate. 

From the PhishStorm dataset, we extract a “blacklist” of 1000 URLs to be labelled as phishing URLs. We use only URL strings without lexical features of contents/URLs to test our privacy-preserving approach against other clustering methods. We then combine this blacklist with a “whitelist” of legitimate domains/URLs from the popular most visited sites by Amazon Alexa. This combined dataset of 2000 URLs is then used to test WPN against other clustering techniques for detection rate on phishing URLs and computation time.

\subsection{Detection Rate}

\begin{figure}[htbp]
\centerline{\includegraphics[width=2.3in]{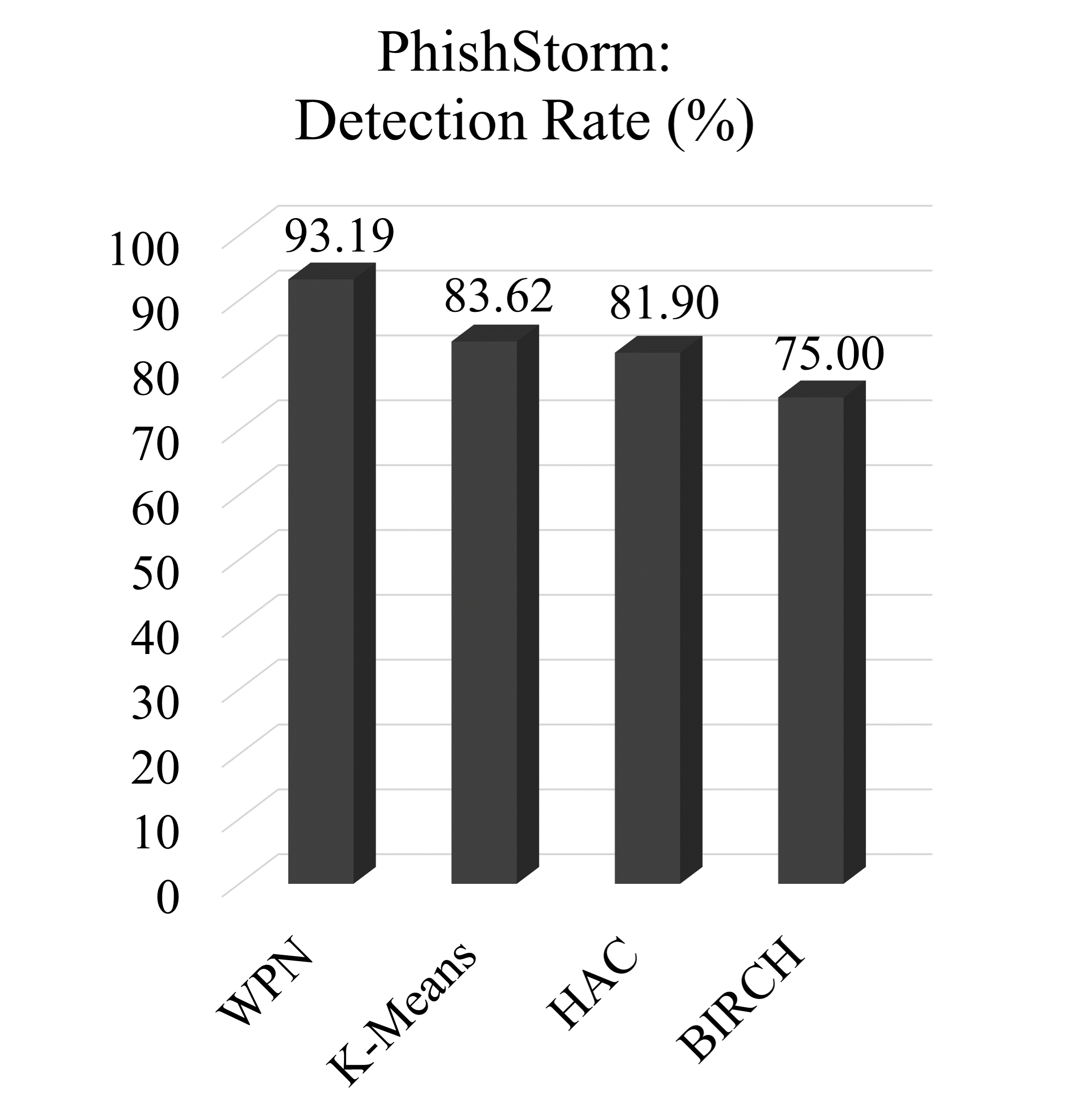}}
\caption{Results of detection rate on PhishStorm data showing comparison of WPN, K-Means, HAC and BIRCH.}
\label{fig4}
\end{figure}

In the first experiment, we run our approach on the aforementioned dataset and compare to popular clustering methods which were initially considered for this system as explained in Section 3 – K-Means, Hierarchical Agglomerative Clustering (HAC) and BIRCH. The results are calculated after from all output clusters except for those containing only white list or only black list URLs as these are inconclusive clusters; clusters containing blacklist URLs with whitelist URLs show that the blacklist URLs are sufficiently similar to the whitelist URLs to claim that they are phishing URLs. The final results are captured as a percentage of phishing domains from our initial blacklist that are captured in the result set by the three algorithms. These are presented in Fig. \ref{fig4}.

\begin{figure}[htbp]
\centerline{\includegraphics[width=3.5in]{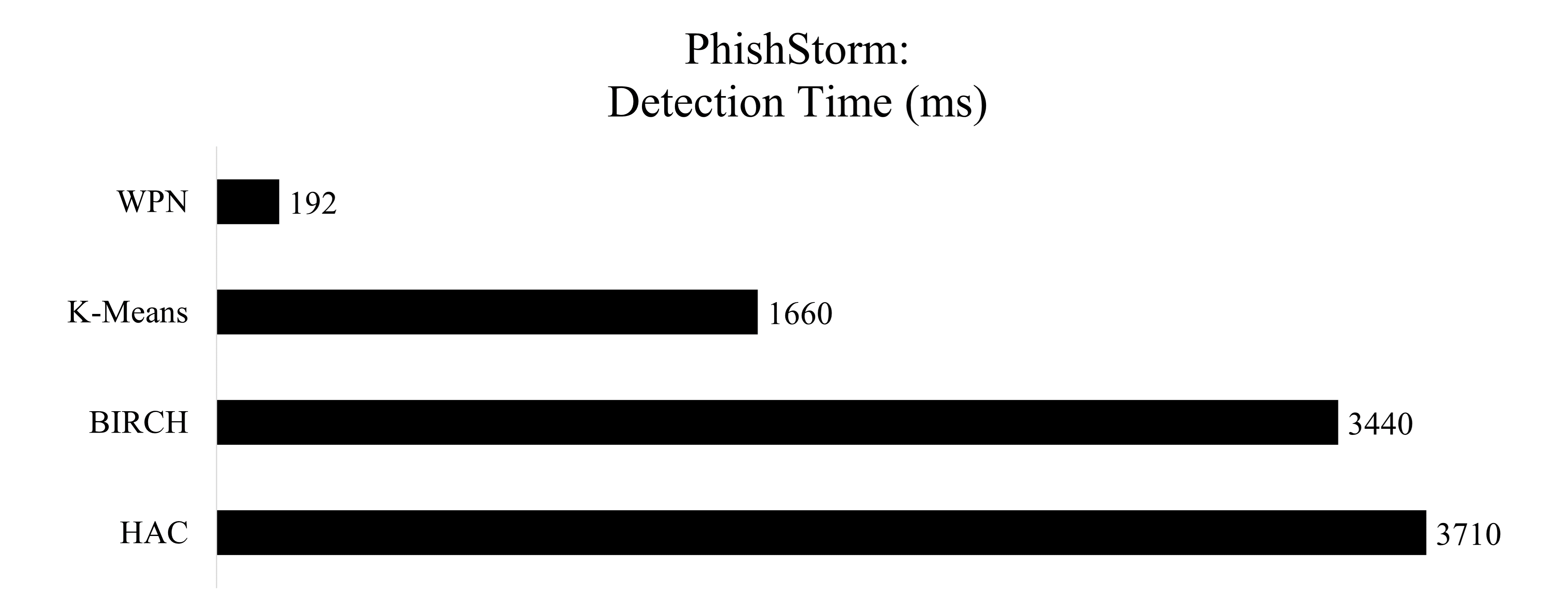}}
\caption{Results of detection time on PhishStorm data showing comparison of WPN, K-Means, HAC and BIRCH.}
\label{fig5}
\end{figure}

Even though K-Means and HAC are able to capture more than 80\% of phishing URLs, the proposed system outperforms both the other unsupervised techniques including K-Means and HAC methods with a 93\% detection rate. We also compare the computation time of these algorithms against WPN and as expected, the proposed system outperforms both other methods in time to process the data and arrive at detection decision; the results are captured in Fig. \ref{fig5} are collected after running the three algorithms on a local machine with Intel i5 chip (8 CPUs) and 16GB RAM.

\subsection{Campaign Detection}

Due to the way data is processed in WPN without pair-wise comparisons, as explained in section 3, the system outputs a set of clusters that enables detection of entire campaign at a time. A sample output of WPN from the previous PhishStorm evaluation is shown in Table \ref{tab2}, highlighting a campaign targeting the popular video game distribution service Steam. The phishing URLs are detected together using the hash-based binning of inputs and grouped together in the same cluster.

\begin{table}[t]
\caption{Campaign detection using WPN.}
\begin{center}
\begin{tabular}{|c|c|c|}
\hline
\textbf{URLs}&\textbf{Cluster}&\textbf{Keywords} \\
\cline{2-3} 
\hline
\textit{steamnation.gegahost.net}&17&steam  \\
\textit{steamlooks.gegahost.net}&17&steam  \\
\textit{steampool.gegahost.net}&17&steam  \\
\textit{steamboom.gegahost.net}&17&steam  \\
\textit{steamjoy.gegahost.net}&17&steam  \\
\hline
\end{tabular}
\label{tab2}
\end{center}
\end{table}

\subsection{Resilience to AI-based Phishing}

Lastly, we tested the WPN system on its ability to deal with evolved phishing attempts using AI-generated phishing domains/URLs. A dataset was constructed using a GPT -3 based open-source large language model (LLM). The GPT-3 is a generative transformer-based LLM trained on 175 billion parameters \cite{b21}. The model was prompted to produce 714 phishing URLs mimicking multiple organizations that are common targets for phishing attempts including those in the banking, retail and health sectors. The results from the model formed the blacklist in this experiment and was combined with 800 whitelist URLs to form the evaluation dataset shown in Table \ref{tab3}. The results are outlined in Table \ref{tab4} below. 

\begin{table}[htbp]
\caption{Dataset - AI-generated Phishing}
\begin{center}
\begin{tabular}{|c|c|}
\hline
\textbf{Legitimate Domains(Whitelist)}&\textbf{Generated URLs (Blacklist)} \\
\cline{1-2}
\hline
\textbf{800}&714 \\
\hline
\end{tabular}
\label{tab3}
\end{center}
\end{table}

\begin{table}[htbp]
\caption{Comparison of Detection Rate (D/R) on AI-generated phishing URLs.}
\begin{center}
\begin{tabular}{|c|c|c|}
\hline
\textbf{Method}&\textbf{Detected Phishing URLs}&\textbf{D/R (\%)} \\
\cline{2-3} 
\hline
\textit{WPN}&699&97.9  \\
\textit{K-Means}&675&94.5  \\
\textit{HAC}&622&87.1  \\
\hline
\end{tabular}
\label{tab4}
\end{center}
\end{table}

Unsupervised techniques are generally better able to deal with evolution of threat as they do not depend on learned signatures from historical data as in the case of supervised methods. This is reflective in the results with all three approaches being able to capture a vast majority of these unseen phishing URLs. WPN, however, outperforms both K-Means and HAC with a 97.9\% of phishing URLs detected. This is because of the more generalizable hash-based clustering, explained in Section \ref{WPN},  and the following refinement stage which allows for more granular approach to arrive at the final result set, grouping unseen phishing URLs with known legitimate domains/URLs with accuracy.

\section{Conclusion \label{conclusion}}

In this paper, we presented a proactive and scalable approach to phishing detection using hash-based clustering that can accurately detect phishing URLs in a privacy-preserving manner with resilience to evolved phishing techniques such as AI-generated domains.

The approach, WPN, addresses the gaps in current detection systems which have significant historical data requirements for training, as in supervised approaches, or high computational requirements affecting system scalability; requirements such as those resulting from the use of pair-wise comparisons in some unsupervised methods. WPN uses URL strings to detect phishing without the need for additional data such as traffic on these URLs; this enables it to detect URLs proactively as soon as they are registered. 

In addition to this, phishing detection systems deployed today typically use features such as those derived from analysis of email content which affects the privacy of end users. WPN is able to detect phishing URLs using only the URL strings without the need to sift through content of phishing vectors such as emails, thereby enabling privacy-preservation. 

Through use of the efficient and scalable hash-based clustering and subsequent refinement using independent similarity measures, WPN is able to detect accurately entire campaigns of phishing URLs at a time as opposed to each individual URL as done in other popularly used techniques through pair-wise comparisons.
 
The results presented in the paper show the high detection rates of the proposed approach on widely used open-source data. We also present results to show resilience against AI-generated phishing URLs, a practice that has become prevalent amongst attackers with the recent gain in popularity of generative AI techniques and LLMs such as Chat GPT. 

Future work will focus on experimenting with a variety of datasets and benchmark against supervised learning-based approaches and with larger volumes of data to capture the temporal analysis of our approach for broader comparison.



\bibliographystyle{IEEEtran}
\bibliography{IEEEabrv,IEEE}

\begin{thebibliography}{10}
\providecommand{\url}[1]{#1}
\csname url@samestyle\endcsname
\providecommand{\newblock}{\relax}
\providecommand{\bibinfo}[2]{#2}
\providecommand{\BIBentrySTDinterwordspacing}{\spaceskip=0pt\relax}
\providecommand{\BIBentryALTinterwordstretchfactor}{4}
\providecommand{\BIBentryALTinterwordspacing}{\spaceskip=\fontdimen2\font plus
\BIBentryALTinterwordstretchfactor\fontdimen3\font minus \fontdimen4\font\relax}
\providecommand{\BIBforeignlanguage}[2]{{%
\expandafter\ifx\csname l@#1\endcsname\relax
\typeout{** WARNING: IEEEtran.bst: No hyphenation pattern has been}%
\typeout{** loaded for the language `#1'. Using the pattern for}%
\typeout{** the default language instead.}%
\else
\language=\csname l@#1\endcsname
\fi
#2}}
\providecommand{\BIBdecl}{\relax}
\BIBdecl

\bibitem{b1}
{Barry Elad}, ``Phishing statistics by types, country and age group,'' \url{https://www.enterpriseappstoday.com/stats/phishing-statistics.html}, 2023.

\bibitem{b2}
N.~T~N, D.~Bakari, and C.~Shukla, ``Business e-mail compromise - techniques and countermeasures,'' in \emph{2021 International Conference on Advance Computing and Innovative Technologies in Engineering (ICACITE)}, 2021, pp. 217--222.

\bibitem{b3}
\BIBentryALTinterwordspacing
T.~R. Soomro and M.~Hussain, ``Social media-related cybercrimes and techniques for their prevention,'' \emph{Applied Computer Systems}, vol.~24, no.~1, pp. 9--17, 2019. [Online]. Available: \url{https://doi.org/10.2478/acss-2019-0002}
\BIBentrySTDinterwordspacing

\bibitem{b4}
{Lance Spitzner}, ``Phishing - it's no longer about malware (or even email),'' \url{https://www.sans.org/blog/phishing-its-no-longer-about-malware-or-even-email}, 2023.

\bibitem{b5}
\BIBentryALTinterwordspacing
S.~S. Roy, P.~Thota, K.~V. Naragam, and S.~Nilizadeh, ``From chatbots to phishbots? -- preventing phishing scams created using chatgpt, google bard and claude,'' 2024. [Online]. Available: \url{https://arxiv.org/abs/2310.19181}
\BIBentrySTDinterwordspacing

\bibitem{b6}
{Egress}, ``New report reveals that nearly three-quarters (71\%) of ai detectors can’t tell if a phishing email has been written by a chatbot,'' \url{https://www.egress.com/newsroom/new-report-reveals-that-nearly-three-quarters-of-ai-detectors-can-t-tell-if-a-phishing-email-has-been-written-by-a-chatbot}, 2023.

\bibitem{b7}
N.~Valiyaveedu, S.~Jamal, R.~Reju, V.~Murali, and N.~K. M, ``Survey and analysis on ai based phishing detection techniques,'' in \emph{2021 International Conference on Communication, Control and Information Sciences (ICCISc)}, vol.~1, 2021, pp. 1--6.

\bibitem{b8}
A.~C. Bahnsen, E.~C. Bohorquez, S.~Villegas, J.~Vargas, and F.~A. González, ``Classifying phishing urls using recurrent neural networks,'' in \emph{2017 APWG Symposium on Electronic Crime Research (eCrime)}, 2017, pp. 1--8.

\bibitem{b9}
\BIBentryALTinterwordspacing
H.~Le, Q.~Pham, D.~Sahoo, and S.~C.~H. Hoi, ``Urlnet: Learning a url representation with deep learning for malicious url detection,'' 2018. [Online]. Available: \url{https://arxiv.org/abs/1802.03162}
\BIBentrySTDinterwordspacing

\bibitem{b11}
\BIBentryALTinterwordspacing
A.~Aljofey, Q.~Jiang, Q.~Qu, M.~Huang, and J.-P. Niyigena, ``An effective phishing detection model based on character level convolutional neural network from url,'' \emph{Electronics}, vol.~9, no.~9, 2020. [Online]. Available: \url{https://www.mdpi.com/2079-9292/9/9/1514}
\BIBentrySTDinterwordspacing

\bibitem{b12}
\BIBentryALTinterwordspacing
A.~Das, S.~Baki, A.~E. Aassal, R.~M. Verma, and A.~Dunbar, ``{SOK:} {A} comprehensive reexamination of phishing research from the security perspective,'' \emph{CoRR}, vol. abs/1911.00953, 2019. [Online]. Available: \url{http://arxiv.org/abs/1911.00953}
\BIBentrySTDinterwordspacing

\bibitem{b13}
\BIBentryALTinterwordspacing
M.~A. Amanullah, R.~A.~A. Habeeb, F.~H. Nasaruddin, A.~Gani, E.~Ahmed, A.~S.~M. Nainar, N.~M. Akim, and M.~Imran, ``Deep learning and big data technologies for iot security,'' \emph{Computer Communications}, vol. 151, pp. 495--517, 2020. [Online]. Available: \url{https://www.sciencedirect.com/science/article/pii/S0140366419315361}
\BIBentrySTDinterwordspacing

\bibitem{b10}
S.~Mondal, D.~Maheshwari, N.~Pai, and A.~Biwalkar, ``A review on detecting phishing urls using clustering algorithms,'' in \emph{2019 International Conference on Advances in Computing, Communication and Control (ICAC3)}, 2019, pp. 1--6.

\bibitem{b22}
\BIBentryALTinterwordspacing
J.~Zamora, M.~Mendoza, and H.~Allende, ``Hashing-based clustering in high dimensional data,'' \emph{Expert Systems with Applications}, vol.~62, pp. 202--211, 2016. [Online]. Available: \url{https://www.sciencedirect.com/science/article/pii/S0957417416302895}
\BIBentrySTDinterwordspacing

\bibitem{b14}
S.~Lloyd, ``Least squares quantization in pcm,'' \emph{IEEE Transactions on Information Theory}, vol.~28, no.~2, pp. 129--137, 1982.

\bibitem{b15}
M.~Ester, H.-P. Kriegel, J.~Sander, and X.~Xu, ``A density-based algorithm for discovering clusters in large spatial databases with noise,'' in \emph{Proceedings of the Second International Conference on Knowledge Discovery and Data Mining}, ser. KDD'96.\hskip 1em plus 0.5em minus 0.4em\relax AAAI Press, 1996, p. 226–231.

\bibitem{b16}
\BIBentryALTinterwordspacing
P.~Indyk and R.~Motwani, ``Approximate nearest neighbors: towards removing the curse of dimensionality,'' in \emph{Proceedings of the Thirtieth Annual ACM Symposium on Theory of Computing}, ser. STOC '98.\hskip 1em plus 0.5em minus 0.4em\relax New York, NY, USA: Association for Computing Machinery, 1998, p. 604–613. [Online]. Available: \url{https://doi.org/10.1145/276698.276876}
\BIBentrySTDinterwordspacing

\bibitem{b17}
\BIBentryALTinterwordspacing
O.~Jafari, P.~Maurya, P.~Nagarkar, K.~M. Islam, and C.~Crushev, ``A survey on locality sensitive hashing algorithms and their applications,'' \emph{CoRR}, vol. abs/2102.08942, 2021. [Online]. Available: \url{https://arxiv.org/abs/2102.08942}
\BIBentrySTDinterwordspacing

\bibitem{b18}
\BIBentryALTinterwordspacing
V.~I. Levenshtein, ``Binary codes capable of correcting deletions, insertions, and reversals,'' \emph{Soviet physics. Doklady}, vol.~10, pp. 707--710, 1965. [Online]. Available: \url{https://api.semanticscholar.org/CorpusID:60827152}
\BIBentrySTDinterwordspacing

\bibitem{b19}
\BIBentryALTinterwordspacing
L.~R. Dice, ``Measures of the amount of ecologic association between species,'' \emph{Ecology}, vol.~26, no.~3, pp. 297--302, 1945. [Online]. Available: \url{http://www.jstor.org/stable/1932409}
\BIBentrySTDinterwordspacing

\bibitem{b20}
S.~Marchal, J.~Francois, R.~State, and T.~Engel, ``Phishstorm: Detecting phishing with streaming analytics,'' \emph{IEEE Transactions on Network and Service Management}, vol.~11, pp. 458--471, 12 2014.

\bibitem{b21}
\BIBentryALTinterwordspacing
T.~B. Brown, B.~Mann, N.~Ryder, M.~Subbiah, J.~Kaplan, P.~Dhariwal, A.~Neelakantan, P.~Shyam, G.~Sastry, A.~Askell, S.~Agarwal, A.~Herbert{-}Voss, G.~Krueger, T.~Henighan, R.~Child, A.~Ramesh, D.~M. Ziegler, J.~Wu, C.~Winter, C.~Hesse, M.~Chen, E.~Sigler, M.~Litwin, S.~Gray, B.~Chess, J.~Clark, C.~Berner, S.~McCandlish, A.~Radford, I.~Sutskever, and D.~Amodei, ``Language models are few-shot learners,'' \emph{CoRR}, vol. abs/2005.14165, 2020. [Online]. Available: \url{https://arxiv.org/abs/2005.14165}
\BIBentrySTDinterwordspacing

\end{thebibliography}

\end{document}